\documentclass[aps,prb,superscriptaddress,reprint]{revtex4-1}

\usepackage{graphicx}
\usepackage{dcolumn}
\usepackage{bm}
\usepackage{setspace}
\usepackage{todonotes}
\usepackage{gensymb}
\usepackage{cleveref}

\newcommand{\hd}{H_{\mathrm{dep}}}
\newcommand{\hv}{H_{\mathrm{vortex}}^{\mathrm{z}}}

\newcommand{\uh}{\mu_0H}

\begin{document}

\title{Reconfigurable magnetic domain wall pinning at the nano-scale using  vortex-generated magnetic fields}

\author{Aaron~C.~H.~Hurst}
\author{Joshua~A.~Izaac}
\author{Fouzia~Altaf}
\affiliation{School of Physics, M013, University of Western Australia, 35 Stirling Hwy, Crawley WA 6009, Australia.}%

\author{Vincent~Baltz}
\affiliation{SPINTEC, UMR 8191 CNRS/INAC-CEA/UJF-Grenoble 1/Grenoble-INP, F-38054 Cedex, France.}%

\author{Peter J.~Metaxas}
\email{peter.metaxas@uwa.edu.au}
\affiliation{School of Physics, M013, University of Western Australia, 35 Stirling Hwy, Crawley WA 6009, Australia.}%

\date{\today}

\begin{abstract} 
We demonstrate that the stray magnetic field generated beneath magnetic vortex cores can be used to generate nano-scale, localized pinning sites for magnetic domain walls in an underlying, perpendicularly magnetized nanostrip. Moreover, the pinning strength can be tuned by toggling the vortex core polarity. Indeed, switching the core polarity so that it is aligned with the magnetization of the expanding domain (rather than against it) can be used to reduce the vortex-mediated wall depinning field by between 40\% and 90\%, depending on the system geometry. Significant reductions in the depinning field  can also be obtained in narrow strips by shifting the core away from the strip's center. 
\end{abstract}

\maketitle


Control of magnetic domain wall (DW) motion is critical in DW device applications such as  (multi-state) memories or logic circuits\cite{Allwood2005,Fukami2009,Jaworowicz2009,Breitkreutz2014,Lequeux2016,CurrivanIncorvia2016}. Stable positions for  domain walls within ferromagnetic strips are often defined by structurally patterning the strip and creating local constrictions or protrusions\cite{Boulle2011}. However, these can be challenging to produce uniformly  and achieving a balance between positional stability (critical for data endurance) and low power domain wall motion (important for low power write operations) can pose issues\cite{Parkin2008}. Less geometrically severe approaches to pinning include the use of electric fields\cite{Bernand-Mantel2013} or local modifications to anisotropy\cite{Franken2011,Franken2011a,Franken2012,Franken2012a,SerranoRamon2013,Franken2013}. Alternatively, one can use  stray magnetic fields generated by  ferromagnetic elements to pin domain walls moving in a neighboring strip or film\cite{Metaxas2009,Hiramatsu2013,Metaxas2013,vanMourik2014}. This approach offers the attractive property of (nonvolatile) reconfigurability since the magnetic state of the pinning-inducing element may be switched, meaning that the pinning strength can be changed, thus enabling the development of controllable gates for DW motion\cite{Franken2013}. 

Previous studies of stray-field-mediated  pinning  have focused primarily on the use of (quasi-)uniformly magnetized, bistable ferromagnetic elements\cite{Metaxas2009,Hiramatsu2013,vanMourik2014,Breitkreutz2014}. Here however, via micromagnetic simulation,  we  demonstrate domain wall pinning in a nanostrip due to overlying ferromagnetic disks that exhibit the magnetic vortex state, an inherently reconfigurable and non-uniform magnetization configuration\cite{Cowburn1999-2,Shinjo2000,Wachowiak2002}. The vortex state consists of an in-plane curling magnetization with an out of plane magnetized nano-scale core which, in the absence of external torques or asymmetries in the disk geometry, will lie in the disk center. Critical to this work is the fact that the  core generates a localized stray magnetic field which is parallel to the core magnetization \cite{Shinjo2000,Wachowiak2002,Rondin2013} and which is shown here to be capable of pinning domain walls in the underlying strip.  Furthermore, the core magnetization direction (perpendicular to the disk plane) can be (dynamically) switched in sign using current or field\cite{Okuno2002,VanWaeyenberge2006,Yamada2007} to toggle between a strong or weak field-mediated DW pinning site. The core can be shifted away from the center of the strip which can also reduce the pinning strength, offering an alternative route for pinning tuning via the controllable vortex core position. Finally, we note that the  core size is largely independent of disk size (for those disks which exhibit a single vortex state: $\approx 100$ nm up to micrometer scale) \cite{Usov1993} making the core a geometrically robust,  naturally tuneable source of localized magnetic field. These results demonstrate that reliable, tailorable DW pinning sites can be generated at the nano-scale without fabricating similarly nano-sized structural features.

\begin{figure}[htbp]
\centering
	\includegraphics[width=6cm]{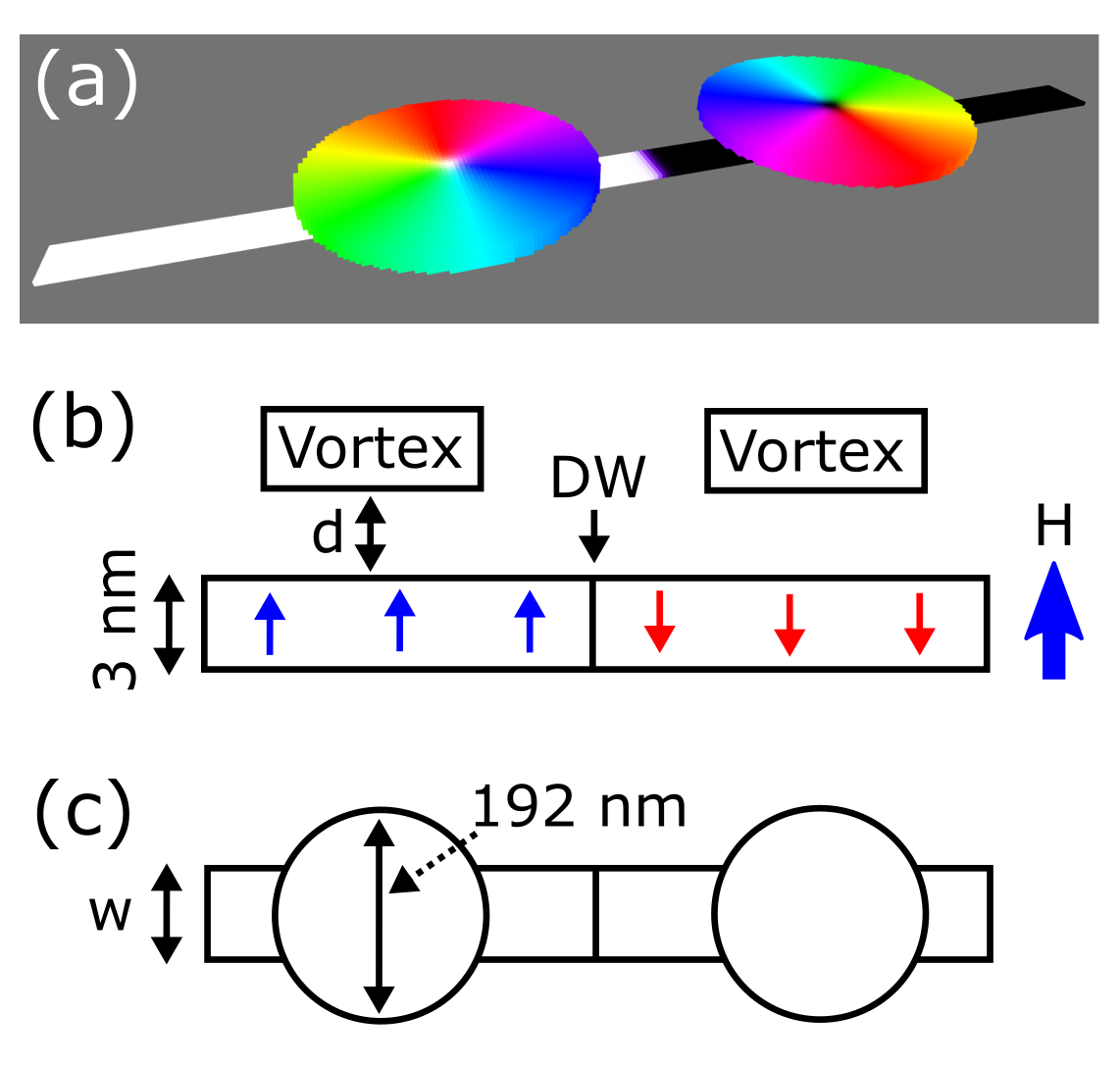}
	\caption{System schematics. (a) 2 vortices above a DW-containing strip in zero external field (visualized using Muview2\cite{muview2}). (b) Side view schematic showing the finite separation, $d$, between the lower surface of the disks and the upper surface of the strip. (c) Top-down schematic with strip width marked as $w$. Schematics in (b,c) are not to scale.}
	\label{systemvis}
\end{figure}

Vortex-mediated DW pinning was studied in a 3 nm thick strip with perpendicular magnetic anisotropy. Two vortex-containing disks with diameters 192 nm and thicknesses 12 nm lie above the strip (Fig.~\ref{systemvis}). There is a vertical separation, $d$, between the bottom of the disks and the top of the strip [Fig.~\ref{systemvis}(b)]. The 768 nm long strip is  CoPtCr-like\cite{Martinez20112} with saturation magnetization $M_S=300$ kA/m, out-of-plane uniaxial anisotropy $K=0.2$ MJ/m$^3$ and exchange stiffness $A_{ex}=10$ pJ/m. The disks are  NiFe-like with $M_S=860$ kA/m, $A_{ex}=13$ pJ/m and negligible intrinsic anisotropy. The damping parameter was 1 everywhere since we consider relaxed states only. No temperature effects were included nor was inter-element exchange coupling meaning that the disk-layer interaction is purely magnetostatic. The disks are symmetrically spaced either side of the center of the strip and along the strip's long axis. The majority of the presented data has been obtained with MuMax3\cite{Vansteenkiste2014} using $3\times 3 \times 3$ nm$^3$ discretization cells. Good consistency  with OOMMF\cite{oommf} results were also found which are mentioned below. For both approaches, the system was initialized with a DW-like transition at the middle of the strip and trial vortex configurations in the disks (the vortex chiralities are anti-clockwise unless otherwise specified). The magnetization was then allowed to relax in zero external field using MuMax3's `minimize' routine  (or OOMMF's `relax' routine). Both routines employ a conjugate gradient method to find the ground state magnetization. From this point the simulation was progressed in a quasi-static manner. An external magnetic field, $H_{ext}=(0,0,H)$ [Fig.~\ref{systemvis}(b)], was applied and increased in steps of $\Delta H$ starting at zero field. The minimize/relax routine was run at each field value, starting from the relaxed configuration obtained at the previous field. The system's relaxed configuration was recorded after each step. Note that $H$ was always positive, driving the DW toward the right hand disk [i.e. expanding the left hand domain; Fig.~\ref{systemvis}(b)].

\begin{figure}[htbp]
\centering
	\includegraphics[width=6.5cm]{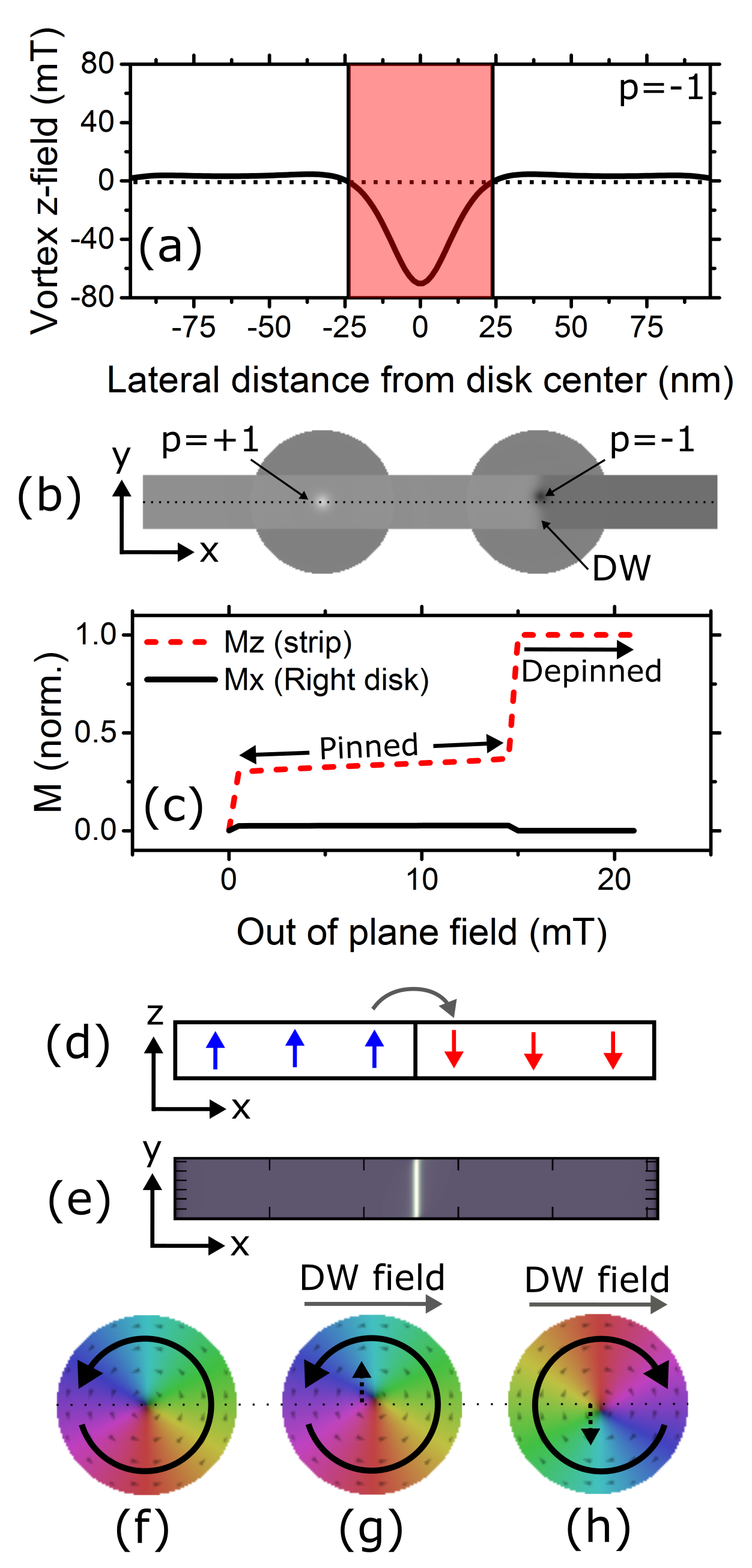}
	\caption{(a) $z-$component of the stray field generated by a $p=-1$ vortex, $\hv$, calculated  10.5 nm below the vortex-containing disc. The (red) shaded region shows where $\hv<0$ and thus opposes $H$, generating DW pinning. (b) Bottom-up visualization of a pinned DW ($w=72$ nm) with gray level mapping of the out-of-plane magnetization component. The DW is at the right hand disk, pinned at the down-magnetized ($p=-1$) vortex core (visible as a dark gray dot). (c) $M_z$ in a $w=36$ nm strip ($d=21$ nm) and $M_x$ in the right hand disk (i.e.~in the disk at which the DW becomes pinned)  as $H$ is increased in steps. (d) An in-plane stray field (arrow) is generated above the DW.  (e) shows the localization of this field (light shading) above the centered DW.  (f) Vortex state with a centered core (i.e.~negligible in-plane field). The circular arrow indicates the chirality. When the DW is pinned at the vortex, the DW's in-plane stray field (gray arrow) shifts the core in the (g) $+y$ direction for an anti-clockwise vortex or (h) the $-y$ direction for a clockwise vortex. The black dotted line across (f-h) marks the position of the unshifted core and the small dotted arrows show the core shift direction.}
	\label{coreshifting}
\end{figure}

We will concentrate first on pinning induced by a vortex which has a negative core polarity, $p=-1$. Directly beneath the $p=-1$ core there is a significant negative out-of-plane field, $\hv$ [Fig.~\ref{coreshifting}(a)]. This  field `peak' has a width (full width half maximum) on the order of 20 nm. and results in there being a localized region in the strip where the negative $\hv$ is strong and  opposes the positive $H$ that is used to drive the DW displacement. This localized field acts as a barrier to DW motion, as can be seen in Fig.~\ref{coreshifting}(b) which shows a snapshot of a DW being pinned to the left of the core ($w=48$ nm). 

In Fig.~\ref{coreshifting}(c), we follow the normalized $z-$component of the magnetization within a 36 nm wide strip ($d=21$ nm) as $H$ is increased by steps, $\Delta H$, enabling us to pass from the pinned state  to the depinned state. At very low external fields ($\uh <1$ mT) the DW moves towards the disk where it is pinned [as in Fig.~\ref{coreshifting}(b)]. At the depinning field, $\hd$,  the wall depins, moves past the vortex and annihilates at the end of the strip. Note that while the wall is pinned at the core, the $x-$component of the magnetization within the vortex increases [also shown in Fig.~\ref{coreshifting}(c)]. This increase corresponds to a shift of the core in the $y$ direction and is driven by the in-plane component  of the stray magnetic field ($+x-$oriented) which exists above the domain wall\cite{Wiebel2005,Wiebel2006,Bellec2010,Metaxas2012} [Figs.~\ref{coreshifting}(d,e)]. This field acts on the vortex's curling magnetization [Fig.~\ref{coreshifting}(f)], inducing a core shift. The direction of this shift depends on the vortex chirality [Figs.~\ref{coreshifting}(g,h)] since the chirality determines the at-equilibrium core displacement for a given in-plane magnetic field \cite{Schneider2001}.

\begin{figure}[htbp]
\centering
	\includegraphics[width=6.5cm]{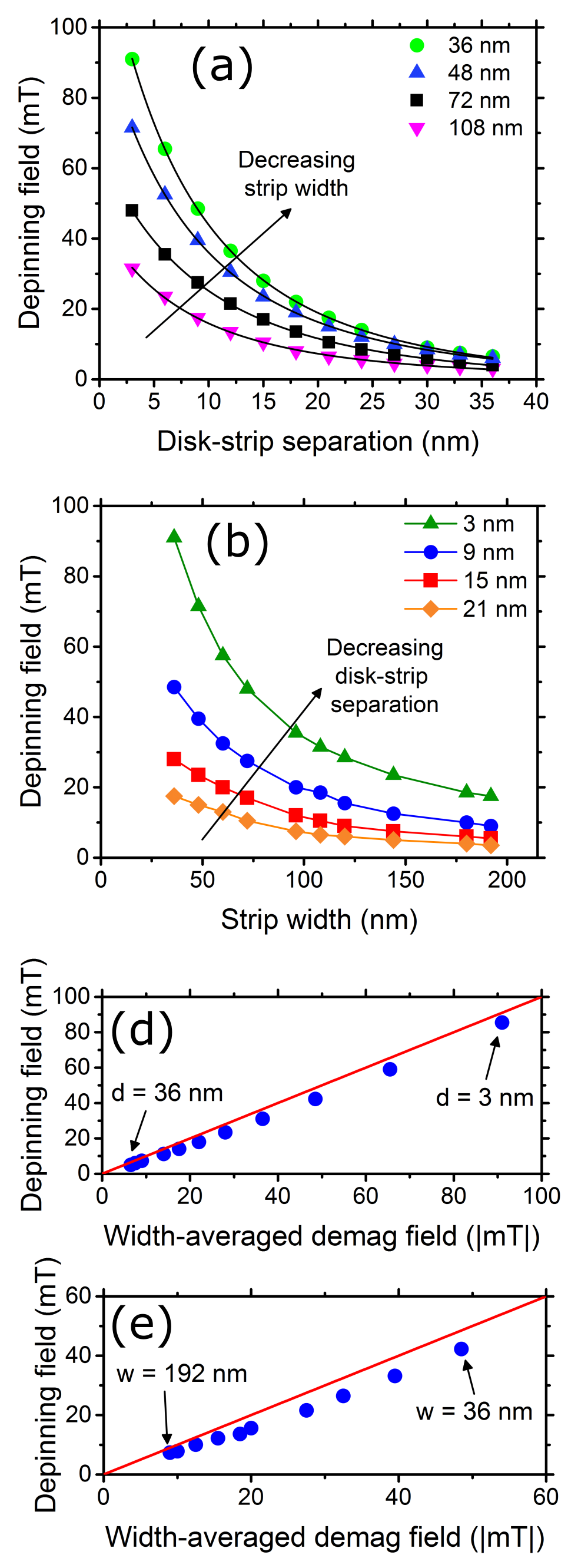}
	\caption{$p=-1$ depinning field  versus (a) disk-strip separation, $d$, and (b) strip-width, $w$, for various values of $w$ and $d$. Black lines in (a) are fits of the form $a + \frac{b}{(c+g)^3}$. Lines in (b) are guides to the eye. Depinning fields for various values of (c)   $d$ ($w=36$ nm) and (d) $w$ ($d=9$ nm) plotted against the absolute value of the $w$-averaged $z-$component of an isolated vortex's stray magnetic field in the region of the DW. The latter was calculated for an isolated disk in the presence of an out of plane field with magnitude $H=\hd-\Delta H$ (i.e.~the field preceding depinning for that particular geometry).}
	\label{depinfields}
\end{figure}

We now look at the dependence of the depinning field on the system geometry for the same $p=-1$ vortex configuration.  In Figs.~\ref{depinfields}(a,b) we show that increasing the strip width, $w$, or the disk-strip separation, $d$, will reduce $\hd$. The latter $d$-dependence occurs because as $d$ is increased, the strip will be further from the disk and thus subject to a weaker $\hv$. Since the pinning is $\hv$-mediated, this leads to a smaller $\hd$. The $w$-dependence can be understood as follows. For very narrow strips, the entire strip width is subject to a strong, negative  $\hv$ [as can be inferred from Fig.~\ref{coreshifting}(a)], which generates a high $\hd$. In contrast, for a wide strip, only the central portion of the strip, will be subject to the strong $\hv$ that exists directly below the core. As such, the average $\hv$ acting across the strip is lower, leading to a reduced $\hd$. Indeed $\hd$ can be shown to closely match the width-averaged $\hv$ both when varying $d$ [Fig.~\ref{depinfields}(d); $w=36$ nm] and $w$ [Fig.~\ref{depinfields}(e);  $d=9$ nm]. Note that $\hv$ was calculated in each case for an isolated vortex subject to $\hd-\Delta H$ (i.e.~at the field preceding that which generated depinning). We also note that the above results were checked with OOMMF\cite{oommf} for d=9 and 15 nm with agreement in the $\hd$ values within 1 mT. Smaller discretization lengths were also examined via MuMax3 as a further check where reducing the in-plane discretization from 3 nm to 2 nm was found to result in a weak change in the de-pinning fields (+2.9\%).

\begin{figure}[htbp]
\centering
	\includegraphics[width=8cm]{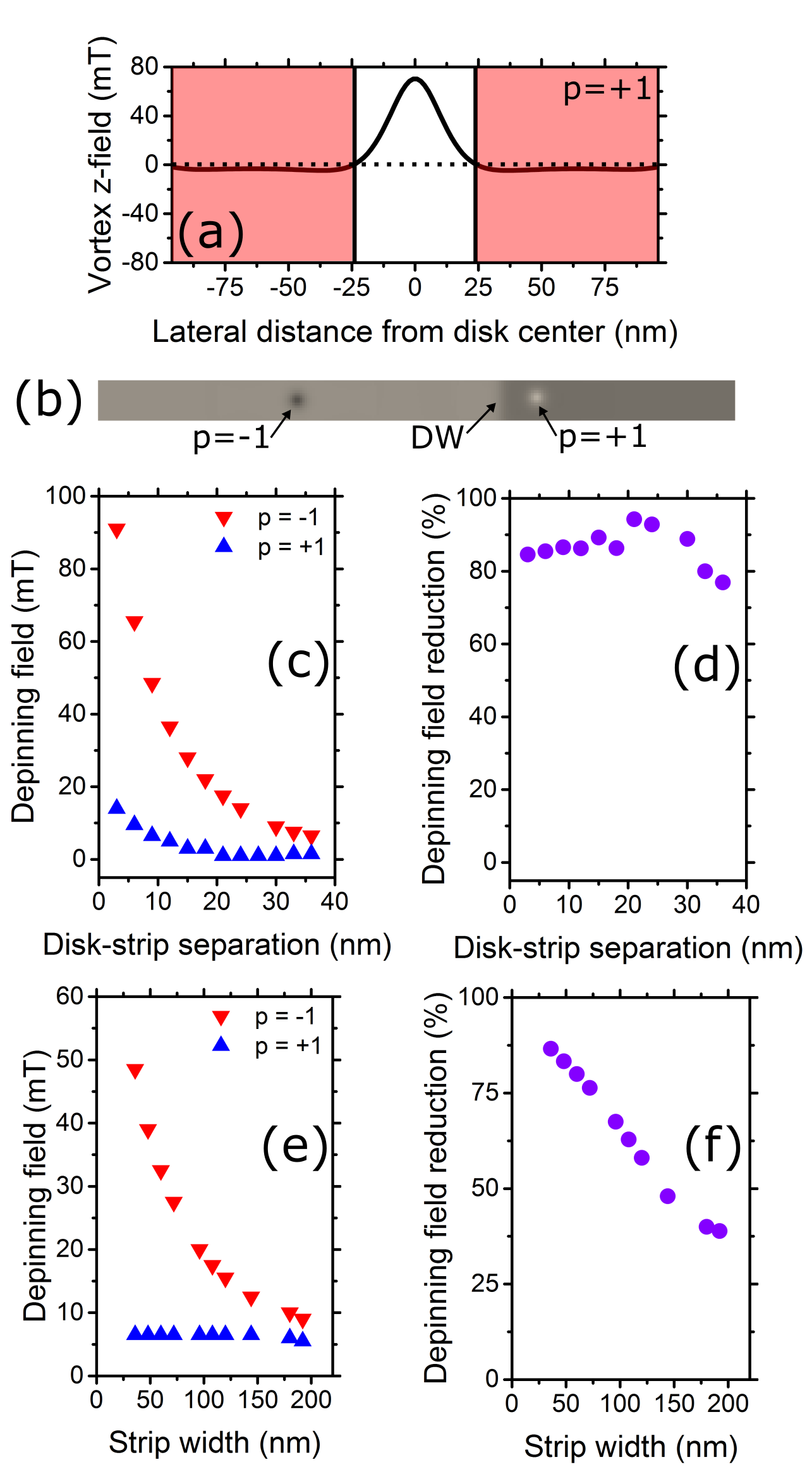}
	\caption{(a) $z-$component of the stray field generated by a $p=+1$ vortex, $\hv$, calculated  10.5 nm below the vortex-containing disc. The (red) shaded region shows where $\hv<0$ and thus opposes $H$, generating DW pinning. (b) Visualization of a pinned DW with gray level mapping of  the out-of-plane magnetization component and superposition of the cores (gray and white dots). The DW is at the right hand disk, pinned well to the left of the right hand vortex core. Lower panels show comparisons of depinning fields for polarities opposing ($p=-1$) and aligned with ($p=+1$) the external field versus (f) $d$ ($w=36$ nm) and (g) $w$ ($d=9$ nm). Corresponding percentage reductions in the depinning fields by switching from $p=+1$ to $p=-1$ are shown in (h,i).}
	\label{depinfields2}
\end{figure}

Switching the polarity of the core to $p=+1$ (achieved by changing the polarity of the pre-relaxation trial vortex state) switches the sign of the vortex stray field, resulting in $\hv$ now being negative (and thus opposing DW motion) in the regions \textit{surrounding} the core position (rather than the region directly beneath the core) [Fig.~\ref{depinfields2}(a); compare to Fig.~\ref{coreshifting}(b) which shows the stray field for $p=-1$].  As a result, the DW now becomes pinned before it reaches the core [Fig.~\ref{depinfields2}(b)], rather than being pinned at the core location [Fig.~\ref{coreshifting}(b)]. The part of the $\hv$ profile causing pinning is also much weaker, resulting in $\hd$ being strongly reduced for $p=+1$. This can be seen in Fig.~\ref{depinfields2}(c) where $\hd$  has been plotted versus $d$ for both core polarities ($w=36$ nm). Switching the core polarity (something which can be achieved rapidly for quick gate modification\cite{VanWaeyenberge2006,Yamada2007}) reduces the depinning field by $\approx 80$ \% over a wide range of $d$ [Fig.~\ref{depinfields2}(d)]. As can be seen in Fig.~\ref{depinfields2}(e) however, $\hd$ for $p=+1$ is almost independent of $w$ (tested for $d=9$ nm). This is consistent with the weak lateral spatial gradient in $\hv$ in the region away from the core [i.e.~the part of the vortex-generated field distribution which generates pinning is relatively uniform in space; Fig.~\ref{depinfields2}(a)]. As a result, the {\%}-reduction in $\hd$ when going from $p=-1$ to $p=+1$ is large only for small $w$ since  $\hd$ for $p=-1$ quickly becomes weak at high $w$ due to the averaging effect discussed above [Fig.~\ref{coreshifting}(e)].

\begin{figure}[htbp]
\centering
	\includegraphics[width=7cm]{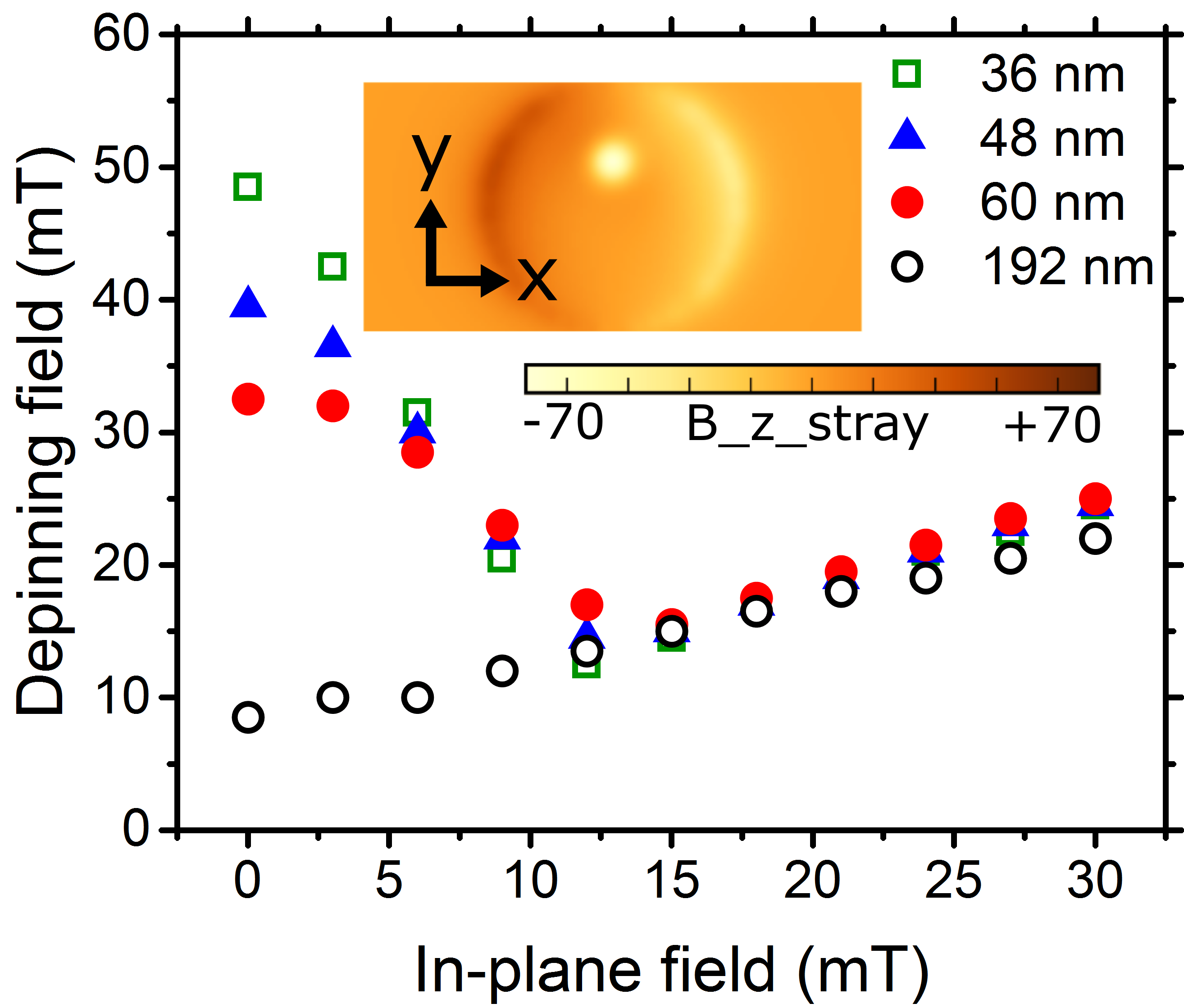}
	\caption{Depinning field versus in-plane field  for four strip widths at $d=9$ nm. The field is applied in the $+x$ direction, shifting the core in the $+y$ direction.  The inset shows the $z$-component of the stray field below a shifted vortex core ($p=-1$) in an isolated disk, as calculated at the vertical center of the strip region for $d=9$ nm in a $+x$-oriented in-plane magnetic field of +21 mT (color scale bar for the out-of-plane field is also shown in mT). The position of the $+y$ shifted core can be identified by its localized negative field.  Stray fields are also generated below the disk's $\pm x$ edges. }
	\label{inplane}
\end{figure}

We now examine the effect that shifting the core has on domain wall depinning (the polarity is fixed here  with $p=-1$. To induce the core shift, we apply an in-plane magnetic field which is constant in time. The field is aligned along the strip's long axis ($+x$),  acting to shift the core perpendicular to that axis ($+y$). $\hd$ versus in-plane field is plotted for four strip widths in Fig.~\ref{inplane}. The inset of Fig.~\ref{inplane} shows a visualization of the out-of-plane disk stray field, calculated at the position of the strip for an in-plane field of 21 mT. From the location of its stray field, the core can be seen to be clearly shifted. 

In terms of the resulting in-plane field dependence of $\hd$, two clear regimes are seen in the three narrowest strips. For low in-plane fields (below 12 mT in Fig.~\ref{inplane}), $\hd$ initially decreases with in-plane field. In this regime, shifting the core away from the narrow strips' centers reduces the average core field that acts on the strip, thereby reducing $\hd$. The initial drop-off in $\hd$ is steepest for the narrower strips since the core-displacement-driven change in the strip-width-averaged $\hv$ for small core displacements will be highest. At higher fields, $\hd$ begins to increase with in-plane field. In this regime there are two separate pinning events occurring. First, the core becomes pinned by the shifted vortex core's field. It then depins, before becoming pinned again at the disk's right edge. This second pinning event arises due to the $+x$ in-plane field transitioning the disk toward a $+x$-magnetized state. This induces   edge magnetic charges on the $\pm x$ sides of the disk which generate out-of-plane stray fields that act on the strip (inset  of Fig.~\ref{inplane}). The out-of-plane components of the edge fields  are negative on the disk's right hand side and are thus capable of pinning the domain wall at that location. Furthermore, the strength of these edge fields grows with in-plane field as the disk acquires a higher  $+x$ magnetization. This pinning mechanism dominates core-mediated pinning, leading to a $\hd$ which increases with in-plane field. Note that compared to the core stray field, these edge fields are relatively uniform in the $y$-direction which results  in similar $\hd$ values for each strip width. Finally we note that $\hd$ in the 192 nm strip exhibits an increasing trend with in-plane field  which is similar to that seen for the narrow strips above 12 mT where depinning requires the wall to move past the edge-charge-generated pinning. Indeed, from 6 mT onwards in the 192 nm strip, pinning due to edge charges dominates the core-mediated pinning with the latter being intrinsically weak for large strip widths  [Fig. 3(b)]. 

In summary, this work demonstrates the potential to use localized stray magnetic fields generated by vortex states to tune domain wall pinning in an underlying perpendicularly magnetized ferromagnetic strip. We evidence that domain wall pinning strength can be reconfigured by switching the vortex core polarity or by modifying the core's lateral position using field-induced vortex manipulation. This approach enables the development of controllable gates for domain wall motion and programmable logic units. These findings also open perspectives for further investigations on harnessing magnetostatic interactions between skyrmions\cite{Sampaio2013b} and domain walls or vortices. Extensions to current-induced manipulation of magnetic textures rather than field-induced may also be envisaged. Beyond the nature of the magnetic texture in use and the way to manipulate them, there will also be interesting effects to examine in full dynamic simulations  where one can aim to (e.g.) rapidly toggle the pinning strength (via a switch in the core magnetization or a well-timed displacement) or dynamically study core displacements (or core-polarity-switching\cite{Wohlhuter2015}) driven by domain-wall-vortex interactions.

\section*{Acknowledgements}

This work was supported by resources provided by the Pawsey Supercomputing Centre with funding from the Australian Government and the Government of Western Australia. Preliminary work was supported by  iVEC through the use of advanced computing resources located at iVEC@UWA. P.J.M.~acknowledges support from the Australian Research Council's Discovery Early Career Researcher Award scheme (DE120100155) and the University of Western Australia (Research Development Award and Early Career Researcher Fellowship Support schemes). Two of the authors were supported  by  internships from the Pawsey Supercomputing Centre (A.C.H.H.) and iVEC (J.A.I.). The authors thank H.~Fangohr, M.~Albert, R.L.~Novak, M.~Kostylev and J.P.~Fried for useful discussions and/or assistance. 

\bibliography{refs_sub}

\end{document}